# Light-emitting diodes by bandstructure engineering in van der Waals heterostructures

F. Withers[1], O. Del Pozo-Zamudio[2], A. Mishchenko[1], A. P. Rooney[3], A. Gholinia[3], K. Watanabe[4], T. Taniguchi[4], S. J. Haigh[3], A. K. Geim[5], A. I. Tartakovskii[2], K. S. Novoselov[1]

[1]School of Physics and Astronomy, University of Manchester, Oxford Road, Manchester, M13 9PL, UK

[2]Department of Physics and Astronomy, University of Sheffield, Sheffield S3 7RH, UK

[3]School of Materials, University of Manchester, Oxford Road, Manchester, M13 9PL, UK

[4]National Institute for Materials Science, 1-1 Namiki, Tsukuba 305-0044, Japan

[5]Manchester Centre for Mesoscience and Nanotechnology, University of Manchester, Oxford Road, Manchester, M13 9PL, UK

**The advent of graphene and related 2D materials[1,2] has recently led to a new technology: heterostructures based on these atomically thin crystals[3]. The paradigm proved itself extremely versatile and led to rapid demonstration of tunnelling diodes with negative differential resistance[4], tunnelling transistors[5], photovoltaic devices[6,7], etc. Here we take the complexity and functionality of such van der Waals heterostructures to the next level by introducing quantum wells (QWs) engineered with one atomic plane precision. We describe light emitting diodes (LEDs) made by stacking up metallic graphene, insulating hexagonal boron nitride (hBN) and various semiconducting monolayers into complex but carefully designed sequences. Our first devices already exhibit extrinsic quantum efficiency of nearly 10% and the emission can be tuned over a wide range of frequencies by appropriately choosing and combining 2D semiconductors (monolayers of transition metal dichalcogenides). By preparing the heterostructures on elastic and transparent substrates, we show that they can also provide the basis for flexible and semi-transparent electronics. The range of functionalities for the demonstrated heterostructures is expected to grow further with increasing the number of available 2D crystals and improving their electronic quality.**

The class of 2D atomic crystals[1], which started with graphene[2] now includes a large variety of materials. However, the real diversity can be achieved if one starts to combine several such crystals in van der Waals heterostructures[3,8]. Most attractive and powerful is the idea of band-structure engineering, where by combining several different 2D crystals one can create a designer potential landscape for electrons to live in. Rendering the band-structure with atomic precision allows tunnel barriers, QWs and other devices, based on the broad choice of 2D materials.

Such band-structure engineering has previously been exploited to create LEDs and lasers based on semiconductor heterostructures grown by molecular beam epitaxy[9]. Here we demonstrate that using graphene as a transparent conductive layer, hBN as tunnel barriers and different transition metal dichalcogenides (TMDC)[1,10] as the materials for QWs, we can create efficient LEDs; Fig. 1F. In our devices, electrons and holes are injected to a layer of TMDC from the two graphene electrodes. Because of the long lifetime of the quasiparticles in the QWs (determined by the height and thickness of the neighbouring hBN barriers), electrons and holes recombine, emitting a photon. The emission wavelength can be fine-tuned by the appropriate selection of TMDC and quantum efficiency (QE) can be enhanced by using multiple QWs (MQWs).

We chose TMDC because of wide choice of such materials and the fact that monolayers of many TMDC are direct band gap semiconductors[11-15]. Until now, electroluminescence (EL) in TMDC devices has been reported only for lateral monolayer devices and attributed to thermally assisted processes arising from impact ionization across a Schottky barrier[16] and formation of p-n junctions[15,17,18]. The use of vertical heterostructures allows us to improve the performance of LED in many respects: reduced contact resistance, higher current densities allowing brighter LEDs, luminescence from the whole device area (Figs. 1E,F) and wider choice of TMDC and their combinations allowed in designing such heterostructures. The same technology can be extended to create other QW-based devices such as indirect excitonic devices[19], LEDs based on several different QWs and lasers.

Figure 1 schematically shows the architecture of single quantum well (SQW) and MQW structures along with optical images of a typical device (Fig. 1E). We utilised a peel/lift Van der Waals technique[20] to produce our devices (see Methods and Supplementary Information for further details on device fabrication). In total we measured more than a dozen of such QW structures comprising single and multiple layers of TMDC flakes from different materials: $MoS_2$, $WS_2$ and $WSe_2$. The yield was 100% with every device showing strong electroluminescence which remains unchanged after months of periodic measurements, which demonstrates the robustness of the technology and materials involved.

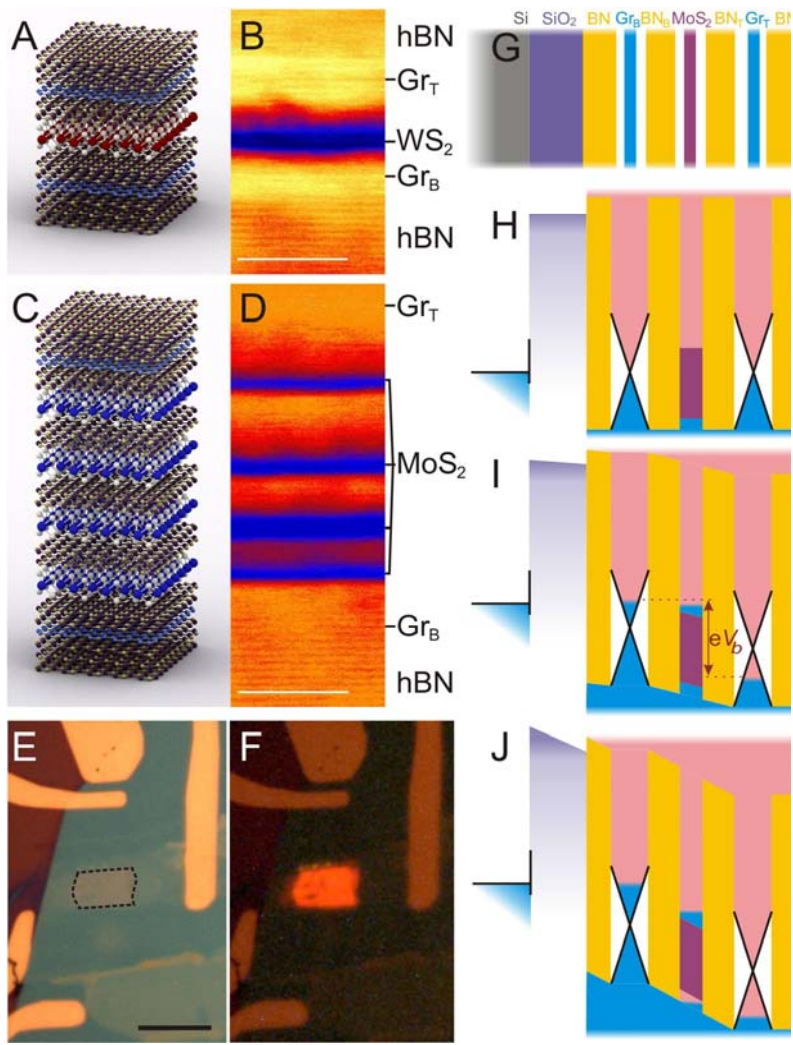

*Fig. 1. Heterostructure devices with SQW and MQW. (**A**) Schematics of SQW heterostructure hBN/Gr$_B$/2hBN/WS$_2$/2hBN/Gr$_T$/hBN. (**B**) Cross-sectional bright field STEM image of the type of heterostructures presented in (A). Scale bar 5nm. (**C, D**) Schematics and STEM imaging of MQW heterostructures hBN/Gr$_B$/2hBN/MoS$_2$/2hBN/MoS$_2$/2hBN/ MoS$_2$/2hBN/MoS$_2$/2hBN/Gr$_T$/hBN. The number of hBN layers between MoS$_2$ QW in (D) varies. Scale bar 5nm. (**E**) Optical image of an operational device (hBN/Gr$_B$/3hBN/MoS$_2$/3hBN/Gr$_T$/hBN). The dashed curve outlines the heterostructure area. Scale bar 10μm. (**F**) Optical image of EL from the same device. V$_b$=2.5V, T=300K. 2hBN and 3hBN stand for bi- and tri-layer h-BN, respectively. (**G**) Schematic of our heterostructure consisting of Si/SiO$_2$/hBN/Gr$_B$/3hBN/MoS$_2$/3hBN/Gr$_T$/hBN. (**H-J**) Band diagrams for the case of zero applied bias (H), intermediate applied bias (I) and high bias (J) for heterostructure presented in (G).*

Cross sectional bright field scanning transmission electron microscope (STEM) images of our SQW and MQW devices demonstrate that the heterostructures are atomically flat and free from interlayer contamination[21]; Fig. 1B,D. The large atomic numbers for TMDC allow the semiconductor crystals to be clearly identified due to strong electron-beam scattering (dark contrast observed in Fig. 1B,D). Other layers were identified by energy dispersive X-ray spectroscopy. The large intensity variation partially obscures the lattice contrast between adjacent layers but, despite this, the hBN lattice fringes can clearly be seen in Figs. 1B, D. The different contrast of the four MoS$_2$ monolayers in the MQW of Fig. 1D is attributed to their different crystallographic orientations, (confirmed by rotating the sample around the heterostructure's vertical direction which changes the relative intensity of different layers).

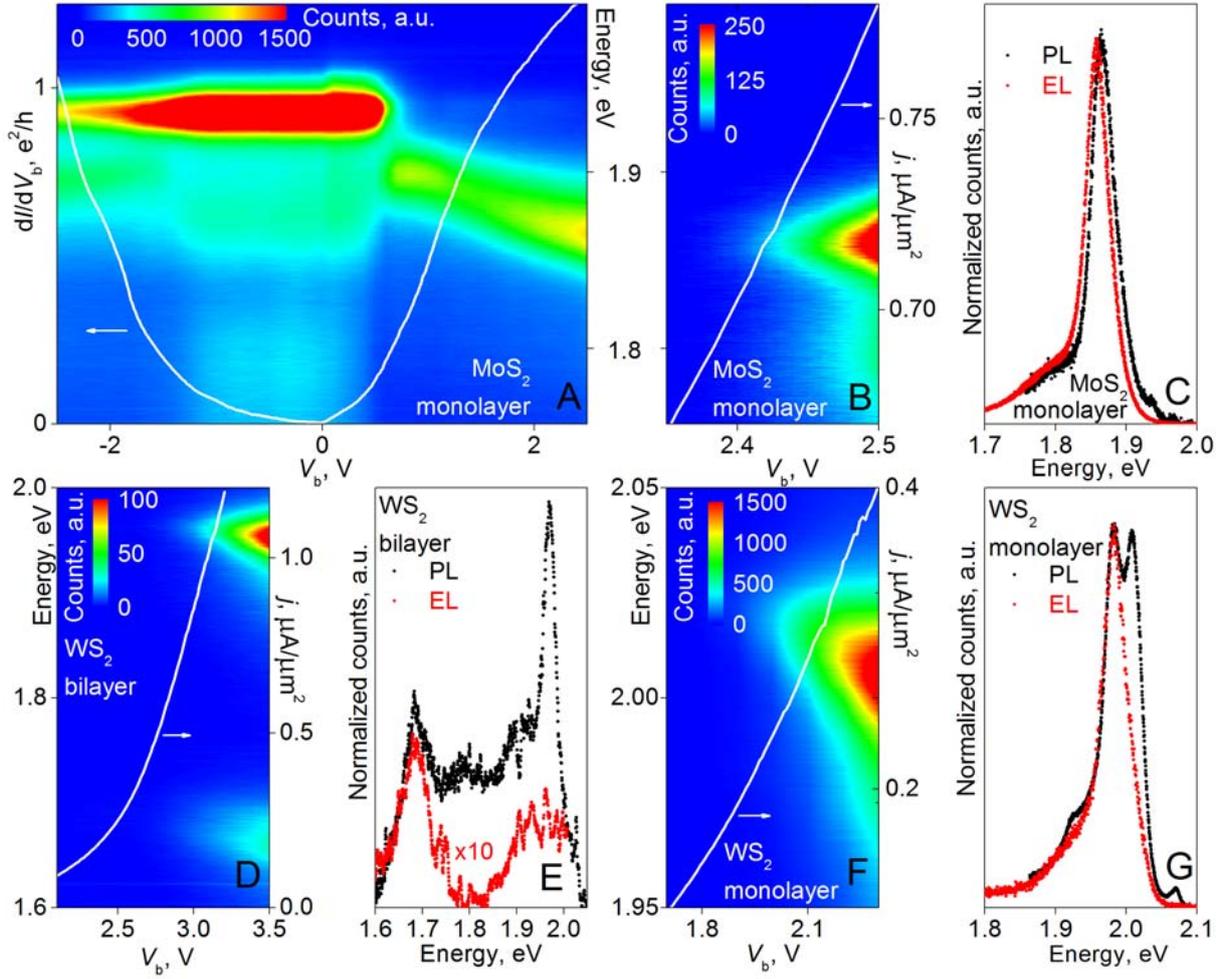

***Fig. 2. Optical and transport characterisation of our SQW devices, T=7K.*** *(A) Colour map of the PL spectra as a function of $V_b$ for a MoS$_2$ –based SQW. The white curve is the device's $dI/dV_b$. Excitation energy $E_L$=2.33eV. (B) EL spectra as a function of $V_b$ for the same device as in (A). White curve: its $j$-$V_b$ characteristics ($j$ is the current density). (C) Comparison of the PL and EL spectra for the same device. As PL and EL occur in the same spectral range, we measured them separately. (D,E) and (F,G) Same as (B,C) but for the bilayer and monolayer WS$_2$ QWs, respectively. The PL curves were taken at $V_b$=2.4V (C), 2.5V (E) and 2V (G); EL – at $V_b$ =2.5V (C), 2.5V (E) and 2.3V (G).*

For brevity we concentrate on current-voltage (*I-V*) characteristics, photoluminescence (PL) and EL spectra from symmetric devices based on MoS$_2$, Fig. 2A-C. Devices based on WS$_2$ and devices with asymmetric barriers are considered in the Supplementary Information.

At low $V_b$, the PL in Fig. 2A is dominated by the neutral A exciton, $X^0$, peak[12] at 1.93 eV. We attribute the two weaker and broader peaks at 1.87 and 1.79 eV to bound excitons[22,23]. At certain $V_b$, the PL spectrum changes abruptly with another peak emerging at 1.90 eV. This transition is correlated with an increase in the differential conductivity (Fig. 2A). We explain this transition as being due to the fact that at this voltage the Fermi level in Gr$_B$ rises above the conduction band in MoS$_2$, allowing injection of electrons into the QW (Fig. 1I). This allows us to determine the band alignment between the Dirac point in graphene and the bottom of conductance band in MoS$_2$: the offset equals half of the bias voltage at which the tunnelling through states in conductance band of MoS$_2$ is first observed. To take into account the effects of possible variance in the thickness of hBN barriers and small intrinsic doping of graphene, we average the onset of tunnelling through MoS$_2$ for positive and negative bias voltages (see Fig. 2A), which yields the offset to be ~0.5eV – in agreement with theoretical prediction[24,25]. Note, that the alignment of graphene's Dirac point with respect to the valence band in hBN has been measured in tunnelling experiments previously[5,26,27].

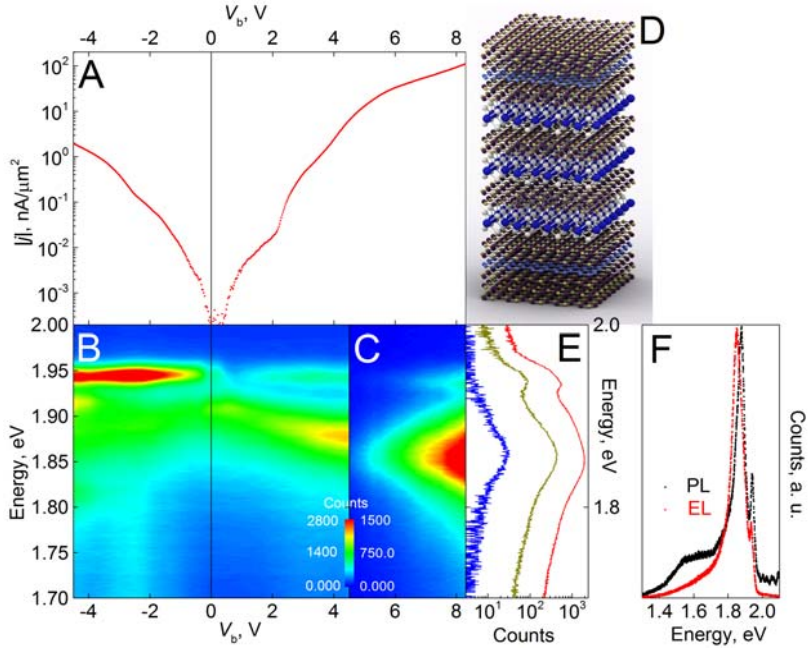

**Fig. 3. Optical and transport characteristics of MQW devices, T=7K.** *(A) Modulus of the current density through a triple QW structure based on MoS$_2$. (B,C) Maps of PL and EL spectra for this device. $E_L$=2.33eV. (D) Its schematic structure. (E) Individual EL spectra plotted in logarithmic scale show the onset of EL at 1.8 nA/μm$^2$ (blue curve). Olive and red: j = 18 and 130 nA/μm$^2$, respectively. (F) Comparison of EL and the PL spectra.*

Injection of electrons into the conduction band of MoS$_2$ leads not only to an increase in tunnelling conductivity but, also, to accumulation of electrons in MoS$_2$ and results in formation of negatively charged excitons[12], X$^-$. The X$^-$ peak is positioned at a lower energy compared to the X$^0$ peak due to the binding energy, $E_B$, of X$^-$. In the case of MoS$_2$ we estimate $E_B$ as ≈36 meV near the onset of X$^-$. As the bias increases, the energy of the X$^-$ shifts to lower values which can be attributed either to the Stark effect or to the increase in the Fermi energy in MoS$_2$[12].

In contrast to PL, EL starts only at $V_b$ above a certain threshold, Figs. 2B. We associate such behaviour with the Fermi level of Gr$_T$ being brought below the edge of the valence band so that holes can be injected to MoS$_2$ from Gr$_T$ (in addition to electrons already injected from Gr$_B$) as sketched in Fig. 1J. This creates conditions for exciton formation inside the QW and their radiative recombination. We find that the EL frequency is close to that of PL at $V_b$≈2.4V (Figs. 2A-C), which allows us to attribute the EL to radiative recombination of X$^-$. Qualitatively similar behaviour is observed for WS$_2$ QWs (see Figs. 2D-G).

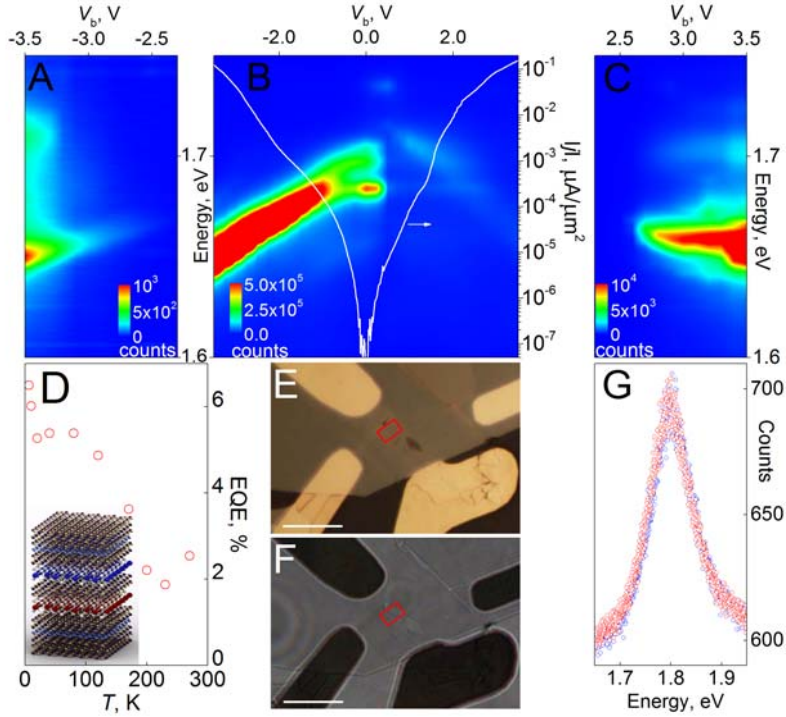

*Fig. 4. Devices combining different QW materials and on flexible substrates. (A-C) EL at negative (A) and positive (C) bias voltages for the device with two QW made from $MoS_2$ and $WSe_2$, schematically shown as the inset in (D). Its PL bias dependence is shown in (B), for laser excitation $E_L$=2.33eV, T=7K. White curve: |j|-$V_b$ characteristics of the device. (D)Temperature dependence of EQE for a device with two QW made from $MoS_2$ and WSe2. Inset: schematic representation of a device with two QW produced from different materials. (E) Optical micrograph taken in reflection of a SQW ($MoS_2$) device on PET. (F) Optical micrograph of the same device as in (E) taken in transmission. For (E-F) the area of the stack is marked by red squares; scale bars are 10μm. (G) EL spectra for the device in (E-F) at zero (blue dots) and 1% (red dots) strain. $V_b$=-2.3V, I=-40μA at room T.*

An important parameter for any light emission device is the QE defined as $\eta = N2e/I$, (here $e$ is the electron charge, $N$ is the number of the emitted photons). For SQW's we obtain quantum efficiencies of ~ 1% - this value by itself is ten times larger than that of planar p-n diodes[15,17,18] and 100 time larger than EL from Schottky barrier devices[16]. Our rough estimations show that the EQE for PL is lower than that for EL. Relatively low EQE found in PL indicates that the crystal quality itself requires improvement and that even higher EQE in EL may then be achieved[28].

To enhance QE even further, we have employed multiple QWs stacked in series, which increases the overall thickness of the tunnel barrier and enhances the probability for injected carriers to recombine radiatively. Fig. 3 shows results for one of such MQW structures with three $MoS_2$ QWs (layer sequence: Si/SiO$_2$/hBN/Gr$_B$/3hBN/$MoS_2$/3hBN/$MoS_2$/3hBN/$MoS_2$/3hBN/Gr$_T$/hBN) and another MQW with four asymmetric $MoS_2$ QWs (Fig. 1C,D) is described in Supplementary Information. The current increases with $V_b$ in a step-like manner, which is attributed to sequential switching of the tunnelling current through individual $MoS_2$ QW's. PL for the MQW device is qualitatively similar to that of SQW devices but the $X^0$ peak is replaced with $X^-$ peak at a $V_b$=0.4V; Fig. 3B. The $X^0$ peak reappears again at $V_b$>1.2V. This can be explained by charge redistribution between different QWs. The EL first becomes observable at $V_b$>3.9V and $j$ of 1.8 nA/μm$^2$; Fig. 3C,E. This current density is nearly 2 orders of magnitude smaller than the threshold current required to see EL in similar SQW. Importantly, the increased probability of radiative recombination is reflected in higher QE, reaching values of ~8.4% (for the device with quadruple QW, 6% for triple). Importantly, this high QE is comparable to the efficiencies of the best modern day organic light emitting devices (OLED's)[29].

The described technology of making designer MQWs offers the possibility of combining various semiconductor QWs in one device. Figs. 4A-C describe an LED made from $WSe_2$ and $MoS_2$ QWs: *$Si/SiO_2/hBN/Gr_B/3hBN/WSe_2/3hBN/MoS_2/3hBN/Gr_T/hBN$*. EL and PL occur here in the low-*E* part of the spectra and can be associated with excitons and charged excitons in $WSe_2$. However, in comparison with SQW devices, the combinational device in Fig. 4 exhibits more than an order of magnitude stronger both PL and EL, ~ 5 % quantum efficiency. We associate this with charge transfer between the $MoS_2$ and $WSe_2$ layers such that electron-hole pairs are created in both layers but transfer to and recombine in the material with the smaller band-gap[30]. Such a process is expected to depend strongly on band alignment, which is controlled by bias and gate voltages. This explains the complex, asymmetric $V_b$ dependence of PL and EL in Fig. 4.

Generally, the fine control over the tunnelling barriers allows reduction in the number of electrons and holes escaping from the quantum well, thus enhancing EQE. EQE generally demonstrates a peak at *T* around 50K-150K, depending on the material. Depending on the particular structure we found that typical values of EQE at room *T* are close or a factor of 2-3 lower than those at low *T*, Fig. 4D.

Finally, we note that because our typical stacks are only 10-40 atoms thick, they are flexible and bendable and, accordingly, can be used for making flexible and semi-transparent devices. To prove this concept experimentally, we have fabricated a $MoS_2$ SQW on a thin PET film; Fig. 4E,F. The device shows PL and EL very similar to those in Fig. 3A-C. We also tested the device's performance under uniaxial strain of up to 1% (using bending) and found no changes in the electroluminescence spectrum; Fig. 4G.

In summary, we have demonstrated band-structure engineering with one atomic layer precision by creating QW heterostructures from various 2D crystals including several TMDC, hBN and graphene. Our LEDs based on a single QW already exhibit quantum efficiency of above 1% and line widths down to 18meV, despite the relatively poor quality of currently available TMDC layers. This EQE can be improved significantly by using multiple QWs, consisting of 3 to 4 QWs, these devices show EQE's up to 8.4%. Combining different 2D semiconductor materials allows fine tuning of the emission spectra and also an enhanced EL with a quantum yield of 5%. These values of quantum efficiency are comparable to modern day OLED lighting and the concept is compatible with the popular idea of flexible and transparent electronics. The rapid progress in technology of CVD growth will allow scaling up of production of such heterostructures.

This work was supported by The Royal Society, Royal Academy of Engineering, U.S. Army, European Science Foundation (ESF) under the EUROCORES Programme EuroGRAPHENE (GOSPEL), European Research Council, EC-FET European Graphene Flagship, Engineering and Physical Sciences Research Council (UK), the Leverhulme Trust (UK), U.S. Office of Naval Research, U.S. Defence Threat Reduction Agency, U.S. Air Force Office of Scientific Research, FP7 ITN S$^3$NANO, SEP-Mexico and CONACYT.

**References:**

1       Novoselov, K. S. *et al.* Two-dimensional atomic crystals. *Proceedings of the National Academy of Sciences of the United States of America* 102, 10451-10453, doi:10.1073/pnas.0502848102 (2005).
2       Novoselov, K. S. *et al.* Electric field effect in atomically thin carbon films. *Science* 306, 666-669, doi:10.1126/science.1102896 (2004).
3       Geim, A. K. & Grigorieva, I. V. Van der Waals heterostructures. *Nature* 499, 419-425, doi:10.1038/nature12385 (2013).
4       Britnell, L. *et al.* Resonant tunnelling and negative differential conductance in graphene transistors. *Nature Communications* 4, 1794, doi:10.1038/ncomms2817 (2013).
5       Britnell, L. *et al.* Field-effect tunneling transistor based on vertical graphene heterostructures. *Science* 335, 947-950, doi:10.1126/science.1218461 (2012).
6       Britnell, L. *et al.* Strong light-matter interactions in heterostructures of atomically thin films. *Science* 340, 1311-1314, doi:10.1126/science.1235547 (2013).
7       Yu, W. J. *et al.* Highly efficient gate-tunable photocurrent generation in vertical heterostructures of layered materials. *Nature Nanotechnology* 8, 952-958, doi:10.1038/nnano.2013.219 (2013).
8       Novoselov, K. S. Nobel Lecture: Graphene: Materials in the Flatland. *Reviews of Modern Physics* 83, 837-849, doi:10.1103/RevModPhys.83.837 (2011).
9       Yao, Y., Hoffman, A. J. & Gmachl, C. F. Mid-infrared quantum cascade lasers. *Nature Photonics* 6, 432-439, doi:10.1038/nphoton.2012.143 (2012).
10      Wang, Q. H., Kalantar-Zadeh, K., Kis, A., Coleman, J. N. & Strano, M. S. Electronics and optoelectronics of two-dimensional transition metal dichalcogenides. *Nature Nanotechnology* 7, 699-712, doi:10.1038/nnano.2012.193 (2012).
11      Mak, K. F., Lee, C., Hone, J., Shan, J. & Heinz, T. F. Atomically thin MoS2: A new direct-gap semiconductor. *Physical Review Letters* 105, 136805, doi:10.1103/PhysRevLett.105.136805 (2010).
12      Mak, K. F. *et al.* Tightly bound trions in monolayer MoS2. *Nature Materials* 12, 207-211, doi:10.1038/nmat3505 (2013).
13      Xiao, D., Liu, G.-B., Feng, W., Xu, X. & Yao, W. Coupled spin and valley physics in monolayers of MoS2 and other group-VI dichalcogenides. *Physical Review Letters* 108, 196802, doi:10.1103/PhysRevLett.108.196802 (2012).
14      Ross, J. S. *et al.* Electrical control of neutral and charged excitons in a monolayer semiconductor. *Nature Communications* 4, 1474, doi:10.1038/ncomms2498 (2013).
15      Ross, J. S. *et al.* Electrically tunable excitonic light-emitting diodes based on monolayer WSe2 p-n junctions. *Nature Nanotechnology* 9, 268-272, doi:10.1038/nnano.2014.26 (2014).
16      Sundaram, R. S. *et al.* Electroluminescence in single layer MoS2. *Nano Letters* 13, 1416-1421, doi:10.1021/nl400516a (2013).
17      Pospischil, A., Furchi, M. M. & Mueller, T. Solar-energy conversion and light emission in an atomic monolayer p-n diode. *Nature Nanotechnology* 9, 257-261, doi:10.1038/nnano.2014.14 (2014).
18      Baugher, B. W. H., Churchill, H. O. H., Yang, Y. & Jarillo-Herrero, P. Optoelectronic devices based on electrically tunable p-n diodes in a monolayer dichalcogenide. *Nature Nanotechnology* 9, 262-267, doi:10.1038/nnano.2014.25 (2014).
19      Rivera, P. *et al.* Observation of long-lived interlayer excitons in monolayer MoSe2-WSe2 heterostructures. *arXiv:1403.4985* (2014).

20  Wang, L. *et al.* One-dimensional electrical contact to a two-dimensional material. *Science* 342, 614-617, doi:10.1126/science.1244358 (2013).
21  Haigh, S. J. *et al.* Cross-sectional imaging of individual layers and buried interfaces of graphene-based heterostructures and superlattices. *Nature Materials* 11, 764–767, doi:10.1038/nmat3386 (2012).
22  Tongay, S. *et al.* Defects activated photoluminescence in two-dimensional semiconductors: interplay between bound, charged, and free excitons. *Scientific Reports* 3, 2657, doi:10.1038/srep02657 (2013).
23  Sercombe, D. *et al.* Optical investigation of the natural electron doping in thin MoS2 films deposited on dielectric substrates. *Scientific Reports* 3, 3489, doi:10.1038/srep03489 (2013).
24  Kang, J., Tongay, S., Zhou, J., Li, J. B. & Wu, J. Q. Band offsets and heterostructures of two-dimensional semiconductors. *Applied Physics Letters* 102, 012111, doi:10.1063/1.4774090 (2013).
25  Sachs, B. *et al.* Doping mechanisms in graphene-MoS2 hybrids. *Applied Physics Letters* 103, 251607, doi:http://dx.doi.org/10.1063/1.4852615 (2013).
26  Lee, G. H. *et al.* Electron tunneling through atomically flat and ultrathin hexagonal boron nitride. *Applied Physics Letters* 99, 243114, doi:10.1063/1.3662043 (2011).
27  Britnell, L. *et al.* Electron tunneling through ultrathin boron nitride crystalline barriers. *Nano Letters* 12, 1707-1710, doi:10.1021/nl3002205 (2012).
28  Gutierrez, H. R. *et al.* Extraordinary room-temperature photoluminescence in triangular WS2 monolayers. *Nano Letters* 13, 3447-3454, doi:10.1021/nl3026357 (2013).
29  Reineke, S. *et al.* White organic light-emitting diodes with fluorescent tube efficiency. *Nature* 459, 234-U116, doi:10.1038/nature08003 (2009).
30  Lee, C. H. *et al.* Atomically thin p-n junctions with van der Waals heterointerfaces. *Nature Nanotechnology* 9, 676-681, doi:10.1038/nnano.2014.150 (2014).

# Light-emitting diodes by band-structure engineering in van der Waals heterostructures

*F. Withers[1], O. Del Pozo Zamudio[2], A. Mishchenko[1], A. P. Rooney[3], A. Gholinia[3], K. Watanabe[4], T. Taniguchi[4], S. J. Haigh[3], A. K. Geim[5], A. I. Tartakovskii[2], K. S. Novoselov[1]*

[1]*School of Physics and Astronomy, University of Manchester, Oxford Road, Manchester, M13 9PL, UK*

[2]*Department of Physics and Astronomy, University of Sheffield, Sheffield S3 7RH, UK*

[3]*School of Materials, University of Manchester, Oxford Road, Manchester, M13 9PL, UK*

[4]*National Institute for Materials Science, 1-1 Namiki, Tsukuba 305-0044, Japan*

[5]*Manchester Centre for Mesoscience and Nanotechnology, University of Manchester, Oxford Road, Manchester, M13 9PL, UK*

# S1. Heterostructure preparation

The quantum well (QW) heterostructures were produced via multiple 'peel' and 'lift' transfer processes as described previously in [1, 2]. Fig. S1 and Fig. S2 show the schematic for the processes. For the 'peel' process a flake is mechanically exfoliated onto a polymer double layer then the bottom polymer is dissolved releasing the membrane which then floats on top of the liquid. The PMMA membrane is then inverted and aligned onto the target crystal. The two crystals are brought into contact and heated until the PMMA adheres to the target substrate. Once the flake has adequately stuck to the target crystal the PMMA membrane is brought back. The flake due to the strong Van der Waals interaction peels from the PMMA onto the target flake.

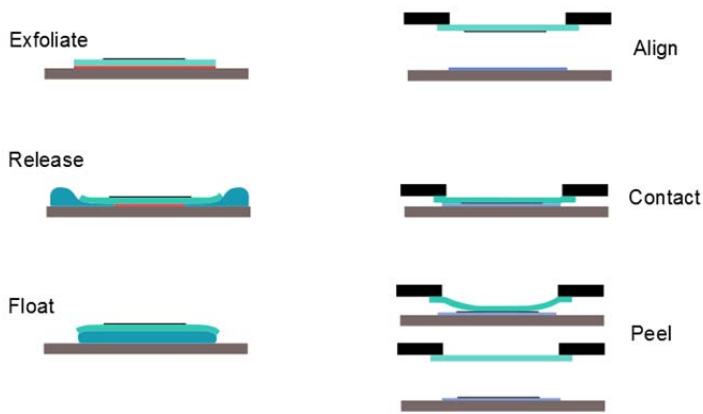

**Fig. S1.** *Schematic procedure for the 'peel' process.*

For the 'lift' process the membrane is produced in the same way as shown in Fig. S1. Instead of peeling the flake onto the target crystal, a large flake on the membrane is used to collect a smaller flake on the substrate. The flake to be lifted is exfoliated onto a second thermally oxidised silicon wafer.

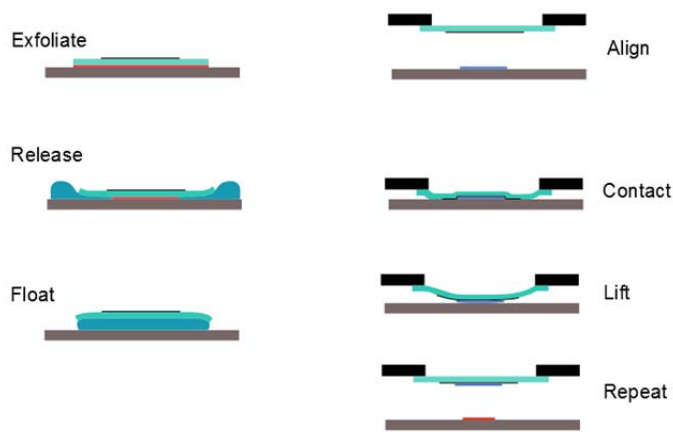

**Fig. S2.** *Schematic procedure for the 'lift' process.*

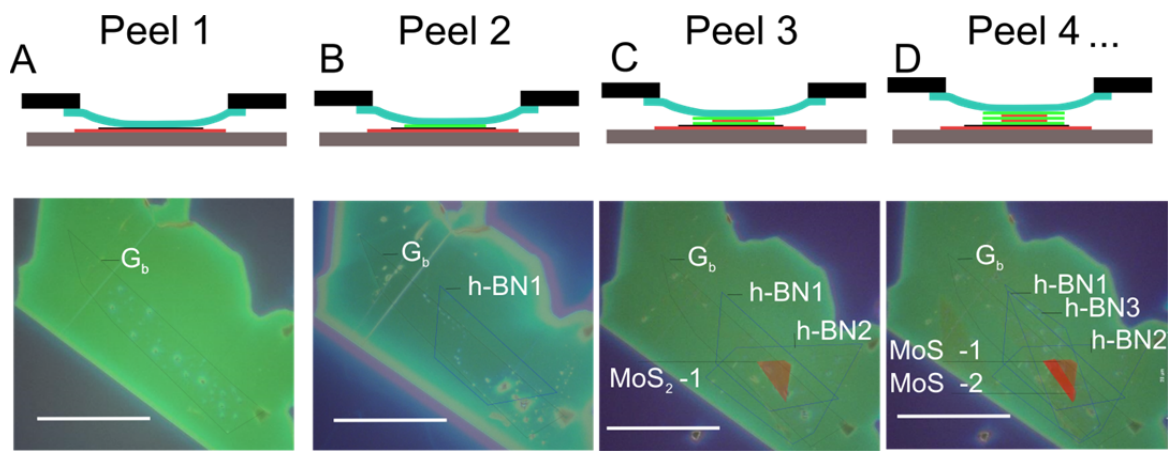

**Fig. S3. (A-D)** *Schematic and differential interference contrast microscope images with semi-transparent dark field images overlaid to highlight flake edges, for the multiple QW structures. Scale bar is 50 μm.*

Fig. S3 shows the fabrication route for the multiple QW (MQW) structure. Firstly a graphene flake is peeled from a PMMA membrane to a hBN crystal on the Si/SiO$_2$ substrate, Fig. S3A. After this a thin h-BN tunnel barrier is peeled from the PMMA membrane onto the hBN-Gr$_B$ structure, Fig. S3B. A thin hBN spacer carrying a single layer TMDC crystal (lifted from a second substrate) is then peeled from the membrane, thus completing the first well, Fig. S3C. This process can be repeated as shown in Fig. S3D to produce a double QW and even further to produce triple and quadruple QW structures.

## S2. Data for additional devices

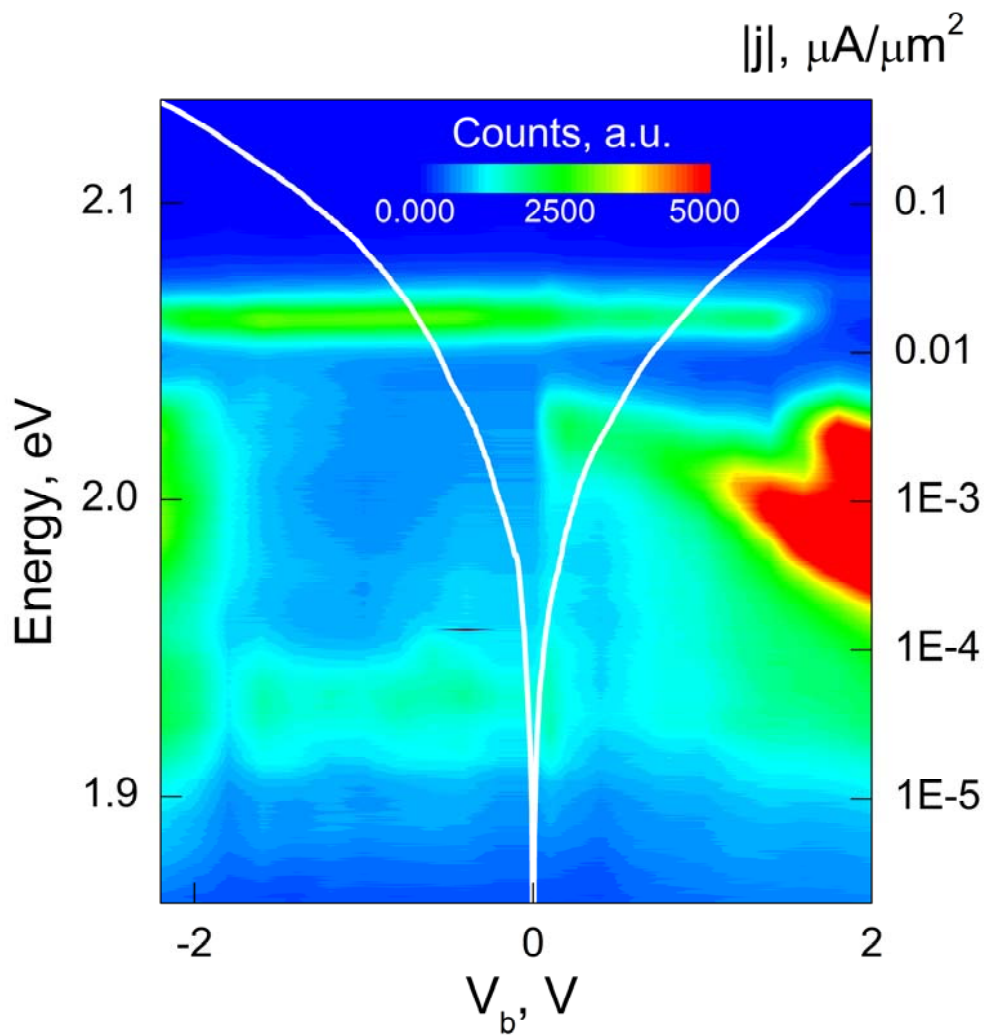

**Fig. S4.** *(A) Colour map of the PL for the WS$_2$ quantum well shown in Figure 2 of the main text $E_L$ = 2.33 eV .*

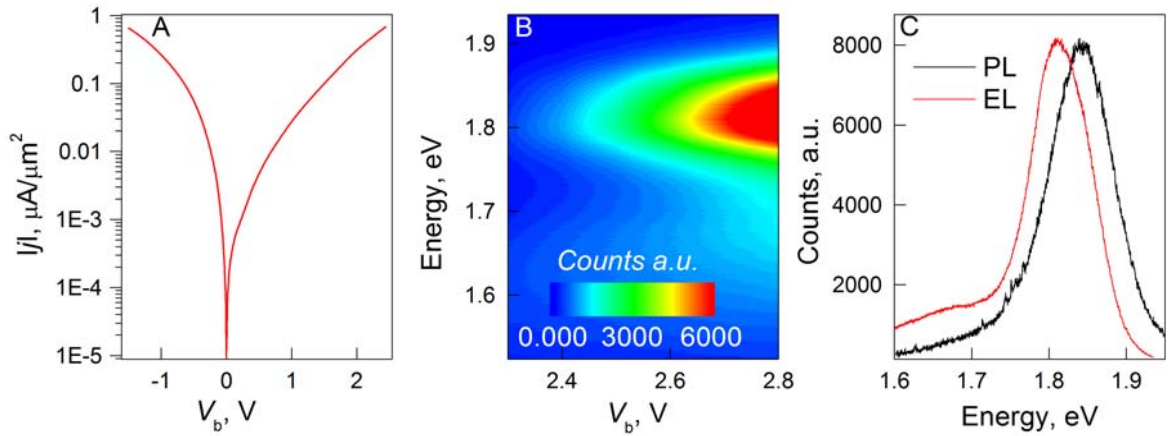

**Fig. S5.** *(A) |j|-$V_b$ characteristics for another SQW made from single layer $MoS_2$. (B) Colour plot of the EL intensity vs bias voltage for the same device. (C) The normalised EL and PL spectra.*

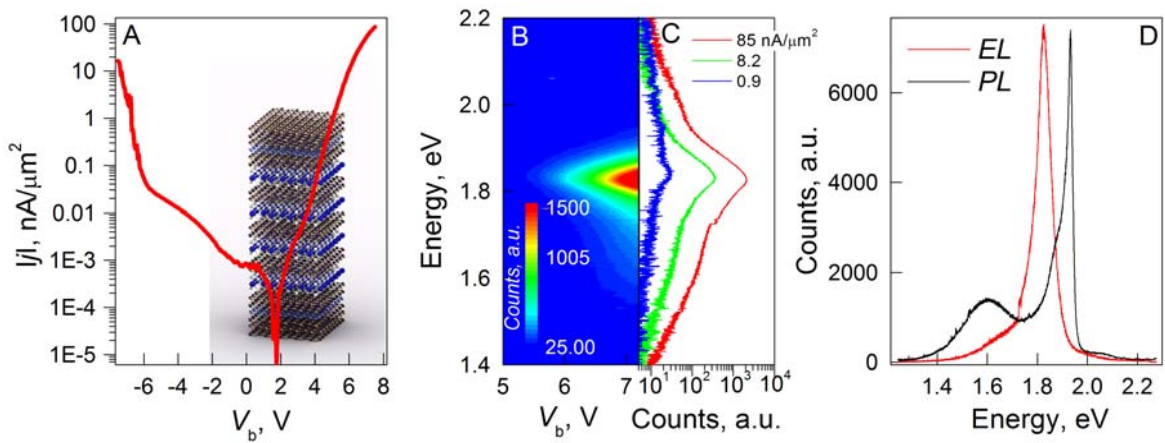

**Fig. S6.** *(A) |j|-$V_b$ characteristics for a device with four QW each made of single layer $MoS_2$. Inset: an image of the approximate structure. (B) Colour map of the EL spectrum. (C) EL spectra for a few different injection currents showing the onset of EL. (D) The EL and PL for the same device (normalised).*

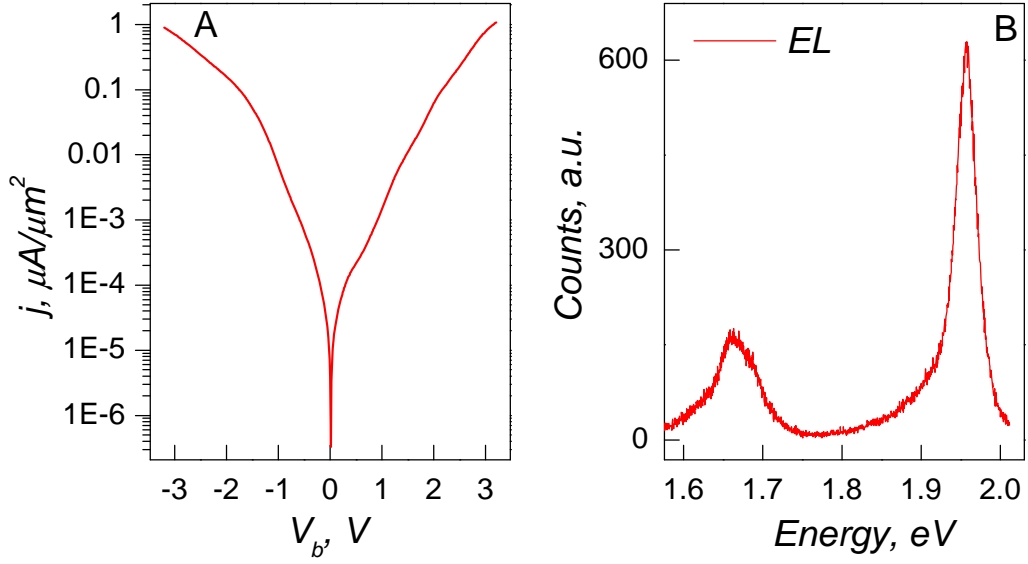

**Fig. S7.** *(A) IjI-$V_b$ characteristics for the bilayer $WS_2$ device described in the main text (B) Single spectrum of the electroluminescence at T = 6 K at a bias voltage of 3.2 V.*

| Device | EL peak position, eV | FWHM, meV | QE at highest injection current and T = 6 K, % |
|---|---|---|---|
| SL $MoS_2$ – 1 | 1.86 | 30 | 0.30 |
| SL $MoS_2$ – 2 | 1.81 | 90 | 0.51 |
| FL $MoS_2$ | 1.86 | 34 | 0.06 |
| SL $WS_2$ | 2.00 | 23 | 1.32 |
| BL $WS_2$ | 1.96 | 29 | 0.09 |
| 3MQW $MoS_2$ | 1.94 | 78 | 6.00 |
| 4MQW $MoS_2$ | 1.83 | 65 | 8.40 |
| 3 stacked $MoS_2$ | 1.82 | 106 | 3.54 |
| $WSe_2$ / $MoS_2$ 2QW-1 | 1.66 | 14 | 4.80 |
| $WSe_2$ / $MoS_2$ 2QW-2 | 1.68 | 21 | 5.40 |

**Table S1.** *The main peak position, width and quantum efficiency for the electroluminescence spectra for different devices based of TMDC materials.*

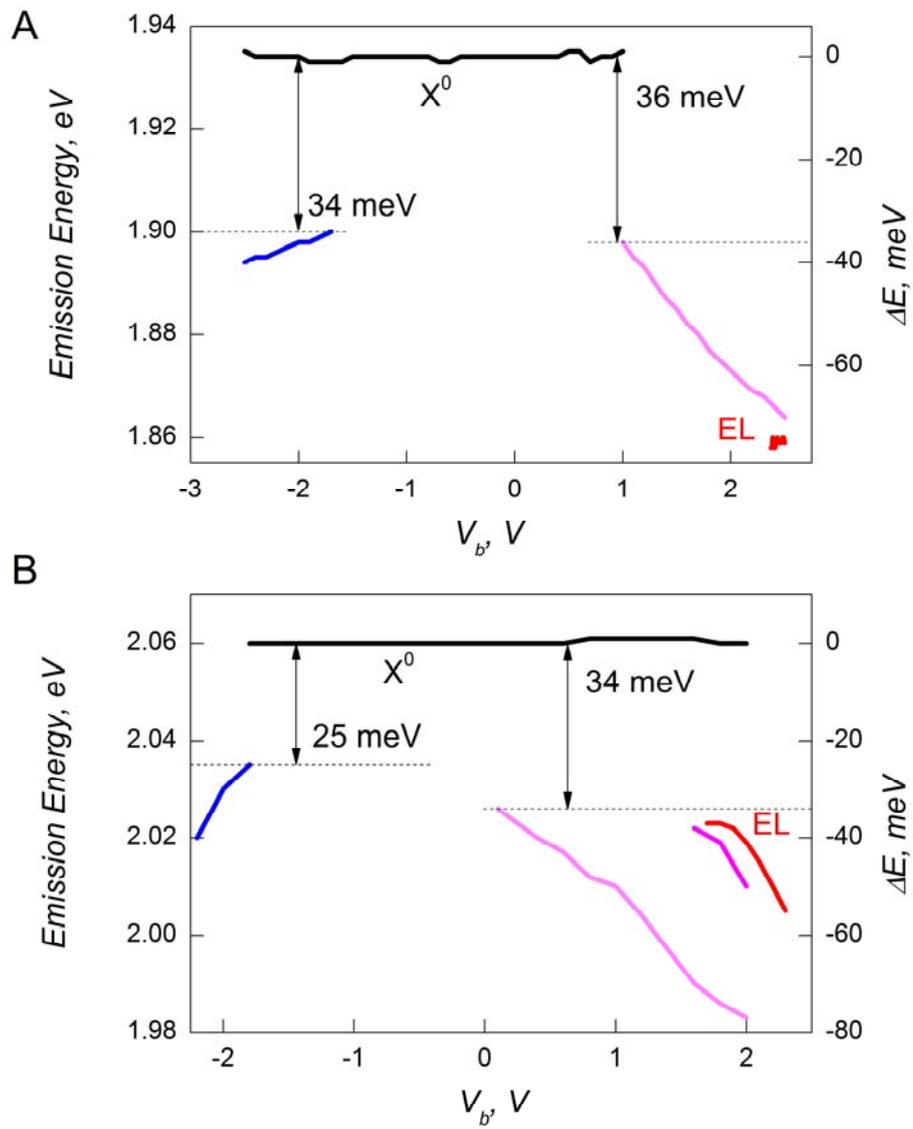

**Fig. S8.** *Extracted peak positions vs bias voltage for **(A)** MoS$_2$, **(B)** WS$_2$.*

# S4. Quantum efficiency

The quantum efficiency is defined as the number of photons emitted per number of injected electron-hole pairs $N2e/i$ ( $N$ = number of emitted photons per second, $e$ electron charge, $i$ is the current passing through a 2 micron area which is our collection area determined by the slit). In order to estimate the number of emitted photons we need to estimate our collection efficiency. The total loss is defined as,

$\eta = \eta_{Lens}\eta_{optic}\eta_{system}$.

$\eta_{optic}$ is the loss of all the optical components in the optical circuit. It was measured directly using a 1.96 eV laser and a power meter to determine the loss at each component. We find $\eta_{optic}$ = 0.18.

$\eta_{system}$ - converts the number of photons arriving at the incoming slit of the detector into the detector counts. It takes into account the loss of photons which pass through the slit, grating and onto the CCD and has been again measured directly by using the 1.96 eV laser and taking spectra of the laser for different powers in order to get a counts vs incident photons. For our system we get 4203 integrated cts/sec per 1 pW. If the system were 100% efficient we should count N=P/hv=3177476 photons, therefore we arrive at an estimate for the system efficiency to be $\eta_{system}$=4203/3177476 =1.32 x $10^{-3}$.

$\eta_{Lens}$ is the efficiency of the lens collection[3]. We use a 100x objective with a numerical aperture, NA = 0.55. The TMDC layer is effectively encapsulated within hBN of refractive index n = 2.2[4].

$$\eta_{Lens} = \frac{1}{4\pi}\int_0^{2\pi} d\varphi \int_0^{\arcsin(\frac{NA}{n})} d\theta \sin\theta = \frac{1}{2}\left[1 - \sqrt{1 - \left(\frac{NA}{n}\right)^2}\right] = 0.016$$

From this we can make an estimate of the quantum efficiency to be, $QE = 2eN_{counts}/\eta i$. In this equation $N_{counts}$ is the integrated number of counts taken for the spectrum.

# S5. Cross sectional imaging

Details of cross sectional imaging of heterostructures produced from 2D materials can be found in [5].

## S5.1 Preparation of cross sectional STEM samples

A dual beam instrument (FEI Dual Beam Nova 600i) has been used for site specific preparation of cross sectional samples suitable for TEM analysis using the lift-out approach [Schaffer, M. *et al.* Sample preparation for atomic-resolution STEM at low voltages by FIB [6]]. This instrument combines a focused ion beam (FIB) and a scanning electron microscope (SEM) column into the same chamber and is also fitted with a gas-injection system to allow local material deposition and material-specific preferential milling to be performed by introducing reactive gases in the vicinity of the electron or ion probe. The electron column delivers the imaging abilities of the SEM and is at the same time less destructive than FIB imaging. SEM imaging of the device prior to milling allows one to identify an area suitable for side view imaging. After sputtering of a 10 nm carbon coating and then a 50 nm Au- Pd coating on the whole surface ex-situ, the Au/Ti contacts on graphene were still visible as raised regions in the secondary electron image. These were used to correctly position and deposit a Pt strap layer on the surface at a chosen location, increasing the metallic layer above the device to ~2 μm. The Pt deposition was initially done with the electron beam at 5kV $e^-$ and 1nA up to about 0.5μm in order to reduce beam damage and subsequently with the ion beam at 30kV $Ga^+$ and 100pA to build up the final 2μm thick deposition. The strap protects the region of interest during milling as well as providing mechanical stability to the cross sectional slice after its removal. Trenches were milled around the strap by using a 30 kV $Ga^+$ beam with a current of 1-6nA, which resulted in a slice of about 1μm thick. Before removing the final edge supporting the milled slice and milling beneath it to free from the substrate, one end of the Pt strap slice was welded to a nanomanipulator needle using further Pt deposition. The cross sectional slice with typical dimensions of 1 μm x 5 μm x 10 μm could then be extracted and transferred to an Omniprobe copper half grid as required for TEM. The slice was then welded onto the grid using Pt deposition so that it could be safely separated from the nanomanipulator by FIB milling. The lamella was further thinned to almost electron beam transparency using a 30kV $Ga^+$ beam and 0.1-1nA. A final gentle polish with Ga+ ions (at 5kV and 50pA) was used to remove side damage and reduce the specimen thickness to 20-70nm. The fact that the cross sectional slice was precisely extracted from the chosen spot was confirmed for all devices by comparing the positions of identifiable features such as Au contacts and /or hydrocarbon bubbles, which are visible both in the SEM images of the original device and within TEM images of the prepared cross section.

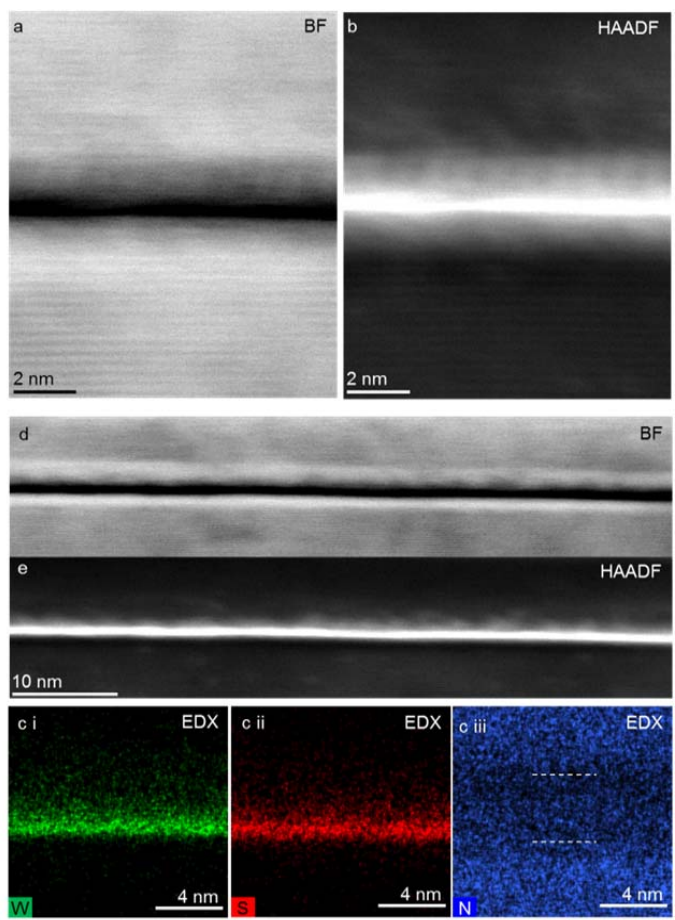

**Fig. S9.** *Bright field (a) and high angle annular dark field (b) STEM images of a single layer WS$_2$ heterostructure cross-section. WS$_2$ lattice fringes are visible in both images, as are the boron nitride fringes at a lower intensity. (c i-iii) Elemental maps for W, S and N extracted from energy dispersive x-ray (EDX) spectrum image data. The top and bottom graphene electrodes can be seen as deficiencies in the nitrogen EDX map (see dashed white lines). This allows the number of boron nitride layers between the graphene and WS$_2$ layer above and below the WS$_2$ layer to be estimated as five and two layers respectively. (d) and (e) show BF and HAADF images at lower magnification.*

## S5.2 Scanning transmission electron microscope imaging and energy dispersive x-ray spectroscopy analysis

High resolution scanning transmission electron microscope (STEM) imaging was performed using a probe side aberration-corrected FEI Titan G2 80-200 kV with an X-FEG electron source operated at 200kV. High angle annular dark field (HAADF) and bright field (BF) STEM imaging was performed using a probe convergence angle of 26 mrad, a HAADF inner angle of 52 mrad and a probe current of ~200 pA. Energy dispersive x-ray (EDX) spectrum imaging was performed in the Titan using a Super-X four silicon drift EDX detector system with a total collection solid angle of 0.7 srad. The multilayer structures were oriented along an <hkl0> crystallographic direction by taking advantage of the Kukuchi bands of the Si substrate.

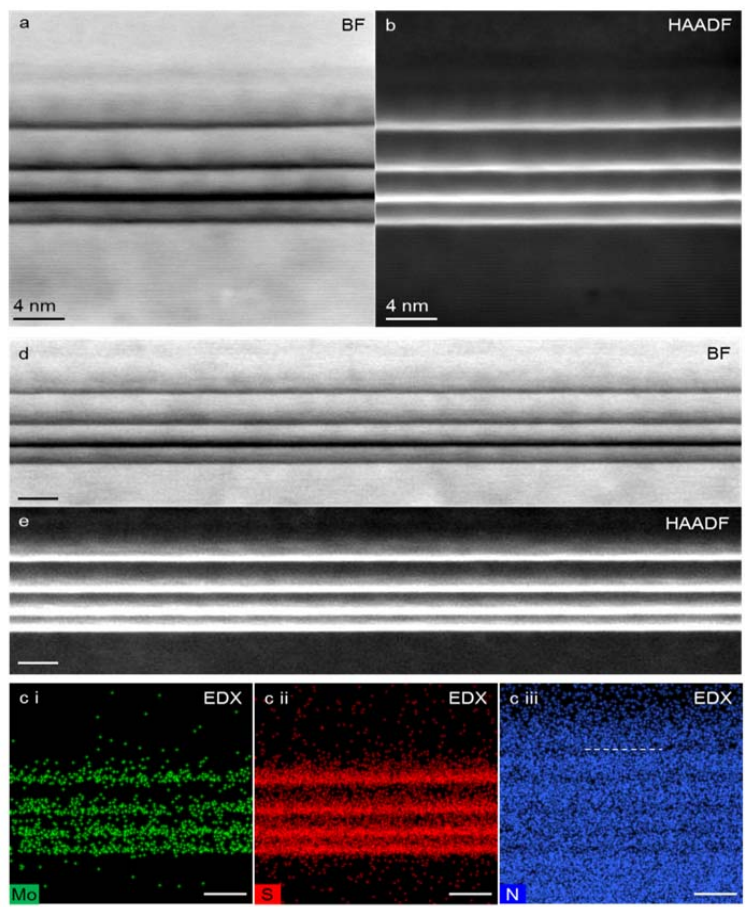

**Fig. S10.** *Cross sectional Imaging of MoS2 multilayer quantum well. (a) Bright field and (b) high angle annular dark field STEM images of the four layer $MoS_2$ heterostructure cross-section. Boron nitride lattice fringes are clearly visible in both images, as are the position of the $MoS_2$ monolayers. The number of boron nitride layers between each $MoS_2$ monolayer were determined to be four, six and nine layers for the bottom, middle and top stacks respectively. (c i-iii) Show elemental maps for Mo, S and N extracted from energy dispersive x-ray (EDX) spectrum image data. The top graphene electrode can be seen as a deficiency in the nitrogen EDX map (indicated by dashed white line in (f)). This allows the number of boron nitride layers between the top $MoS_2$ layer and the graphene to be estimated as seven layers. (d) and (e) show BF and HAADF images at lower magnification. The structure was found to be atomically flat and highly uniform over regions >100nm. All scale bars are 4 nm.*

**References**

1. Wang, L., et al., *One-Dimensional Electrical Contact to a Two-Dimensional Material.* Science, 2013. **342**(6158): p. 614-617.
2. Kretinin, A.V., et al., *Electronic quality of graphene on different atomically flat substrates.* arXiv:1403.5225, 2014.
3. Mueller, T., et al., *Efficient narrow-band light emission from a single carbon nanotube p-n diode.* Nature Nanotechnology, 2010. **5**(1): p. 27-31.
4. Gorbachev, R.V., et al., *Hunting for monolayer boron nitride: optical and Raman signatures.* Small, 2011. **7**(4): p. 465-468.
5. Haigh, S.J., et al., *Cross-sectional imaging of individual layers and buried interfaces of graphene-based heterostructures and superlattices.* Nature Materials, 2012. **11**(9): p. 764–767.
6. Schaffer, M., B. Schaffer, and Q. Ramasse, *Sample preparation for atomic-resolution STEM at low voltages by FIB.* Ultramicroscopy, 2012. **114**: p. 62-71.